\documentclass{elsart}

\usepackage{graphicx}

\usepackage{amssymb}

\begin{document}

\begin{frontmatter}

\title{Agent-based simulation of a financial market}

\author[DIBE]{Marco Raberto\thanksref{email}}
\author[DIBE]{Silvano Cincotti}
\author[Intertek]{Sergio M. Focardi}
\author[DIEE]{Michele Marchesi}

\thanks[email]{E-mail: raberto@dibe.unige.it}

\address[DIBE]{Dipartimento di Ingegneria Biofisica ed Elettronica, Universit\`a di Genova, Via Opera Pia 11a,
16145 Genova, Italy}
\address[Intertek]{The Intertek Group, rue de Javel, 75015 Paris, France}
\address[DIEE]{Dipartimento di Ingegneria Elettrica ed Elettronica, Universit\`a di Cagliari, Piazza d'Armi, 09123 Cagliari, Italy}

\begin{abstract}
This paper introduces an agent-based artificial financial market in which heterogeneous agents trade one single asset through a realistic trading mechanism for price formation. Agents are initially endowed with a finite amount of cash and a given finite portfolio of assets. There is no money-creation process; the total available cash is conserved in time. In each period, agents make random buy and sell decisions that are constrained by available resources, subject to clustering, and dependent on the volatility of previous periods. The model herein proposed is able to reproduce the leptokurtic shape of the probability density of log price returns and the clustering of volatility. Implemented using extreme programming and object-oriented technology, the simulator is a flexible computational experimental facility that can find applications in both academic and industrial research projects.
\end{abstract}

\end{frontmatter}

\section{Introduction and motivation}
Over the last ten years, a number of computer-simulated, artificial financial markets have been built; LeBaron \cite{LeBaron00} offers a review of recent work in this field. Following the pioneering work done at the Santa Fe Institute \cite{PAHLT94,LAP99}, a number of researchers have proposed artificial markets populated with heterogeneous agents endowed with learning and optimisation capabilities. Generally speaking, these markets exhibit behaviour close to that of real world markets but are often too complex to be studied analytically.

Others have proposed artificial markets populated with heterogeneous agents characterised by simple trading rules \cite{CMZ97,BPS97,ST98,LM99,CS99,Iori99,CB00,LLS00}. These lend themselves better to analytical modelling while still retaining the ability to capture fundamental features of market behaviour. 
It is probably fair to say that no artificial market is yet able to explain all the known stylised facts on asset price behaviour. Given the complexity of the task, a compromise between simple generalisation and a faithful representation of realistic detail is called for.

Our objective was to build an artificial market, which we will refer to as the Genoa market, that exhibits realistic trading features and takes into account the finiteness of agents' resources. The goal was to build a robust simulated multi-agent market model on which it would be possible to perform computational experiments using various types of artificial agents.

This paper outlines the framework of the Genoa artificial market and gives a detailed description of its structure. We then present simulation results and close with some remarks on future research directions.

\section{The model's framework}

Following the general scientific principle of conceptual parsimony, our objective is to offer a simple understanding of the known stylised facts of financial time series, i.e., volatility clustering and fat tails in the distribution of short-term returns.

The price-formation process of our market is built around a mechanism for matching demand and supply, the key ingredient of the model. We make the realistic assumption that agents' resources are limited. We assume that, within the trading horizon under consideration (i.e., a few minutes to several days), the global amount of cash in the economy is time-invariant. Money creation is thus explicitly ruled out. Agents are restricted to trading only one asset in exchange for cash.

The adoption of agents with limited resources creates serious constraints on the agent decision-making process. We ran numerous simulations in which agents endowed with a limited amount of cash were divided into subpopulations, adopting either chartist, fundamentalist or random trading strategies. In all these simulations, one population invariably prevailed and the others decayed, losing wealth and relevance. This observation is consistent with what Friedman noted in 1953 \cite{Friedman53}. One possible way out of this situation is to allow random migrations between populations. As it is unlikely an agent embrace a losing strategy, the Genoa market keeps things simple and assumes that orders are randomly issued.

A population of agents that issue random orders in a limited resource market produces a price process with a Gaussian distribution and mean-reverting behaviour, but neither fat tails nor volatility clustering. To represent real world price processes, mechanisms able to better reproduce the behaviour of real traders are needed.

We modelled herding phenomena and a link between market volatility and agent uncertainty. Drawing from Cont and Bouchaud \cite{CB00}, agent aggregation was modelled as clustering in a random graph. The link between nervous (i.e. volatile) markets and agent uncertainty is introduced through the ordering mechanism. In volatile markets, agent uncertainty on asset market prices grows. To represent this, orders are issued at random, but their limit price exhibits a functional dependence on past price volatility. These mechanisms result in a market price process exhibiting fat tails, zero autocorrelation of returns and the serial autocorrelation of volatility.

\section{The Genoa market microstructure}
\label{sec:marketMicrostructure}

Let $N$ be the number of traders and let us denote the $i$-th trader with the subscript $i$. We let time evolve in discrete steps. We denote with $C_{i}(h)$ the amount of cash and with $A_{i}(h)$ the amount of assets owned by the $i$-th trader at time $h$. We denote with $p(h)$ the price of the stock at time $h$. 

At each simulation step, each trader issues a buy order with probability $P_i$ or a sell order with probability $1-P_i\,$. The figures below are relative to simulations where $P_i$ has been set to 0.5 for all agents. 

Suppose the $i$-th trader issues a sell order at time $h+1$. Let's denote with $a_{i}^{s}$ the quantity of stocks offered for sale by the $i$-trader at time $h+1$. We stipulate that the quantity of stocks offered for sale at time step $h+1$ is a random fraction of the quantity of stocks owned at time step $h$ according to the rule: $a_{i}^{s} = [r_{i} \cdot A_{i}(h)]$ where $r_{i}$ is a random number drawn from a uniform distribution in the interval $[0,\;1]$ and the symbol $[x]$ denotes the integer part of $x$. In addition, a limit sell price $s_{i}$ is associated to each sell order. 

We stipulate that sell orders cannot be executed at prices below the limit price. Limit prices are computed as follows: $s_{i} = p(h) / N_{i}(\mu,\sigma_i)$ where $ N_{i}(\mu,\sigma_i)$ is a random draw from a Gaussian distribution with average $\mu = 1.01$ and standard deviation $\sigma_i$. 

It is worth noting that, as $\mu=1.01$, the mean value of all $b_i$ is likely to be greater than $p(k)$, while the mean value of all $s_i$ is likely to be smaller than $p(k)$. In other words, we introduce a spread between the average value of buy/sell orders to represent the fact that a trader placing an order wants to increase the chance of the order being executed. Hence, for a buy order the trader is likely to be willing to pay more than $p(k)$; conversely, for a sell order, the trader is likely to offer the stock at a lower price than $p(k)$.
 
The value of $\sigma_i$ is proportional to the historical volatility $\sigma(T_i)$ of the price $p(h)$ through the equation $\sigma_i = k \cdot \sigma(T_i)$, where $k$ is a constant and $\sigma(T_i)$ is the standard deviation of log-price returns \cite{RSCR99}, calculated in the time window $T_i$. The time window $T_i$ can be different for each trader. Tuning the system by performing numerous simulations with different parameters, we found that appropriate values for $k$ and $T_i$ are: $k = 3.5$ and $T_i = 20$. Graphics in figures 1-3 refer to a simulation run where $k$ has been set to $3.5$ and $T_i$ is 20 time steps long for each trader. 
Linking limit orders to volatility takes into account a realistic aspect of trading psychology: when volatility is high, uncertainty as to the ``true" price of a stock grows and traders place orders with a broader distribution of limit prices.

Buy orders are generated in a fairly symmetrical way with respect to sell orders. If the $i$-th trader issues a buy order at time $h+1$, the amount of cash employed in the buy order, $c_i$, is a random fraction of his or her available cash at time $h$; $c_{i} = r_{i} \cdot C_{i}(h)$, where $r_i$ is a random draw from a uniform distribution in the interval $[0,\;1]$. A limit price $b_{i}$ is associated to each buy order. We stipulate that orders cannot be executed at prices higher than the limit price. 

Limit prices are computed as follows: $b_{i} = p(h) \cdot N_{i}(\mu,\sigma_i)$, where $ N_{i}(\mu,\sigma_i)$ is a random draw from a Gaussian distribution with average $\mu$ and standard deviation $\sigma_i$. As for sell orders, $\mu = 1.01$ and $\sigma_i = k \cdot \sigma(T_i)$, where $k = 3.5$ and $T_i = 20$ for each trader. The quantity of assets ordered to buy $a_{i}^{b}$ is therefore given by $ a_{i}^{b} = [c_{i} / b_{i}]$, where $[x]$ denotes the integer part of $x$.   

It is worth noting that the random numbers $r_i$ and $N_i(\mu,\sigma_i)$ are generated independently at each time step and are different for each trader. Hence traders exhibit heterogeneous random behaviour subject to two constraints: the historical price volatility (which is included in $\sigma_i$) and the finiteness of the resources available to each trader, i.e., $C_i(h)$ and $A_i(h)$.

The price formation process is set at the intersection of the demand and supply curves. We compute the two curves at the time step $h+1$ as follows. Suppose that at time $h+1$ traders have issued $U$ buy orders and $V$ sell orders. For each buy order, let the pair $(a_{u}^{b}, b_u), u=1, 2, ..., U$ indicate respectively the quantity of stocks to buy and the associated limit price. For each sell order in the same time step, let the pair $(a_{v}^{s},s_{v}), v=1, 2, ...,V$ denote respectively the quantity of stocks to sell and the associated limit price. Let us define the functions: 

\begin{equation}
\label{eq:curvaDiAcquisto}
f_{h+1}(p) = \sum_{u\,|\,b_u \ge p}\;a_{u}^{b}\,;
\end{equation}		

\begin{equation}
\label{eq:curvaDiVendita}
g_{h+1}(p) = \sum_{v\,|\,s_{v} \le p}\;a_{v}^{s}\,;
\end{equation}

$f_{h+1}(p)$ represents the total amount of stocks that would be bought at price $p$ (demand curve). It is a decreasing step function of $p$, i.e., the bigger $p$, the fewer the buy orders that can be satisfied. If $p$ is greater than the maximum value of $b_u, u=1, 2, ..., U$, then $f_{h+1}(p) = 0$. If $p$ is lower than the minimum value of $b_{u}, u=1, 2, ..., U$, then $f_{h+1}(p)$ is the sum of all stocks to buy. Conversely, $g_{h+1}(p)$ represents the total amount of stocks that would be sold at price $p$ (supply curve) and is an increasing step function of $p$. In particular, its properties are symmetric with respect to those of $f_{h+1}(p)$.

The clearing price computed by the system is the price $p^*$ at which the two functions cross. We define the new market price at time step $h+1$, $p(h+1)$ as: $p(h+1) = p^*\,$.
The aggregate quantity $f(p^*)$ is number of stocks for which there is a demand at a limit price higher than or equal to $p^*$. The aggregate quantity $g(p^*)$ is the number of stocks offered at a limit price lower than or equal to $p^*$. 
As $f(p)$ and $g(p)$ are step functions, generally the following relation holds: $f(p^*)\neq g(p^*)$.
Hence, in order to keep the total number of stocks unchanged, if $f(p^*) < g(p^*)$ \Big($f(p^*) > g(p^*)$\Big), we stipulate that only $f(p^*)$ \Big($g(p^*)$\Big) stocks are traded. If $f(p^*) < g(p^*)$, then $g(p^*) - f(p^*)$ stocks offered for sale at a compatible limit price are randomly chosen and discarded from the corresponding sell orders. Conversely, if $f(p^*) > g(p^*)$, then $f(p^*) - g(p^*)$ stocks demanded to buy at a compatible limit price are randomly chosen and discarded from the corresponding buy orders.
Buy and sell orders with limit prices compatible with $p^*$ have now the same aggregate quantity and can be executed.  
Following transactions, traders' cash and portfolios are updated. 
Orders that do not match the clearing price are discarded.

It should be noted that, because $f(p)$ and $g(p)$ are step functions, two rare pathological cases may occur. In the first case, $f(p)$ and $g(p)$ do not cross at a single point but have a common horizontal segment whose abscissas are $p^{*1}$ and $p^{*2}$. In this case, we assume: $p^{*}:=(p^{*1}+p^{*2})/2\,$.
In the second case, $f(p)$ and $g(p)$ do not cross at all, i.e., $f(p_1)=0$ and $g(p_2)=0$ as $p_1 < p_2$. In this case, the time step is discarded and a new iteration begins.

At the beginning of the simulation, the current price $p(0)$ is set in an exogenous way and each trader is endowed with a certain amount of cash and a certain amount of stocks. These amounts can be the same for all traders, or may differ.

Lastly, we modelled opinion propagation among agents. We draw on Cont and Bouchaud \cite{CB00} for the application of random graph theory to trading 
networks. As explained in \cite{CB00}, the cluster size distribution follows 
an inverse power law. At the beginning of the simulation, the probability 
$P_i$ of placing a buy order is set at 0.5 for each trader. This probability 
is subsequently updated in function of clustering effects as explained below. 
At each time step, pairs of traders are randomly chosen with probability 
$P_a$. If a pair is chosen, a cluster is formed between the two traders forming the pair. In this way, clusters of traders are progressively formed, grow, and 
eventually merge. 

At each simulation step, a random draw either activates one 
cluster with probability $P_c$ or leaves all clusters inactive with 
probability $1-P_c$. If the system decides to activate a cluster, one cluster is 
then randomly chosen. In this case, all traders belonging to the selected 
cluster change their value of $P_i$ from 0.5 to 1.0 or 0.0 with probability 
50 \%. That is to say, all traders belonging to the selected cluster behave in 
the same way, placing buy orders if $P_i = 1.0$ or sell orders if 
$P_i = 0.0$. After orders are placed, the cluster is destroyed and for each trader belonging to the cluster, values of $P_i$ are set at 0.5. 

Note that, given the finite amount of resources available to traders in the Genoa model, orders are not directly proportional to the respective cluster size. 
The Genoa model introduces an additional correction, making limit orders dependent on volatility. The model thereby introduces price volatility correlations.      

Let's make two further observations. First, though the amount of cash and the number of stocks in the market are time-invariant, the global wealth is a time-varying quantity which is obviously a linear function of the stock price. The price-formation system therefore creates or destroys wealth. A second observation is that the finite resource conditions of the Genoa market induce mechanisms of reversion to the mean for the price process. Suppose that the stock price increases above the equilibrium level at which cash and the global value of stocks are equal. Under the previous assumption, if the aggregate value of stocks exceeds the amount of cash, there will be an imbalance between buy and sell orders that will tend to make prices revert to the equilibrium value.

There are other feedback effects, notably a mechanism that will tend to flatten the distribution of stocks and cash among agents: a trader with a surplus of stocks will tend to issue more sell orders than buy orders and viceversa.

\section{Simulation results}
We will now present some features of a typical simulation 10000 time steps long with: $N = 100$, $P_a = 0.0002$ and $P_c = 0.1\,$.
The initial price of the stock has been fixed at \$ 100  and every trader is endowed with the same value in stock and cash, i.e.: \$ 30000  and a portfolio of 300 stocks.

\begin{figure}
\begin{center}
\includegraphics[scale=0.7]{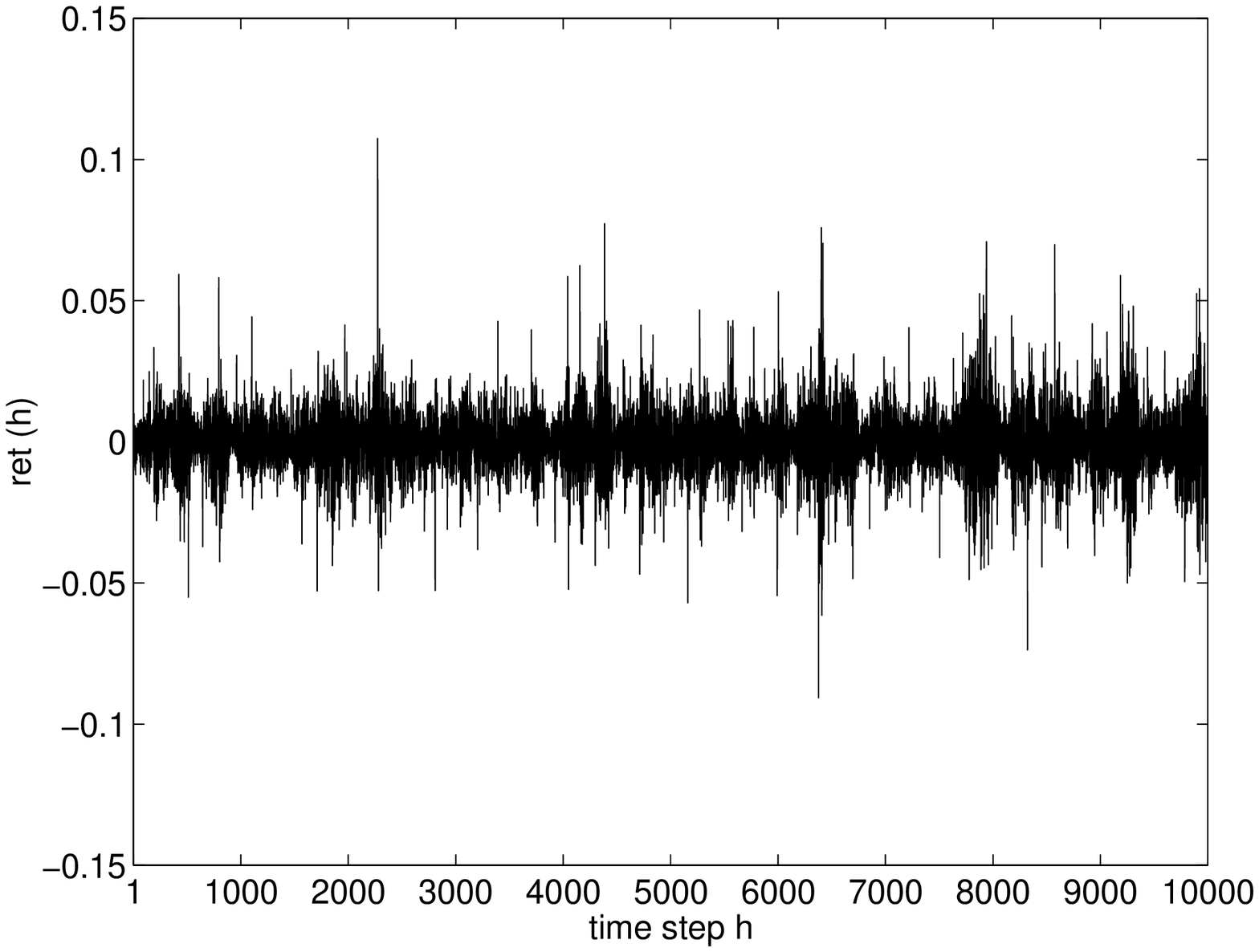}
\end{center}
\caption{Plot of logarithmic returns $ret(h)$ of price $p(h)$: $ret(h) = \log p(h+1) - \log p(h)\,$.}
\label{ret}
\end{figure}

\begin{figure}
\begin{center}
\includegraphics[scale=0.7]{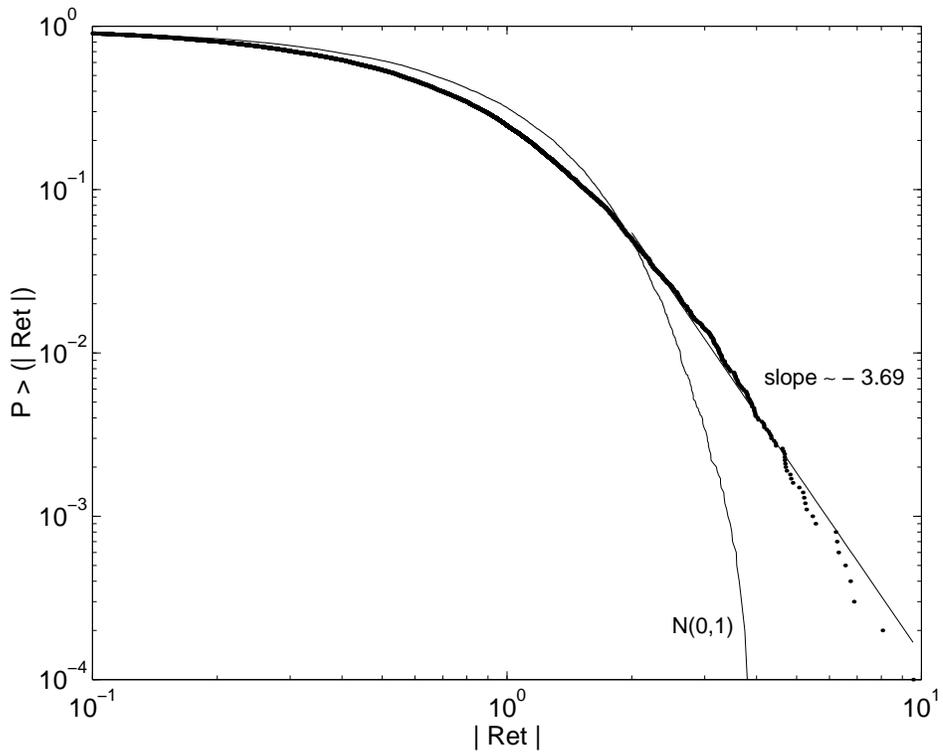}
\end{center}
\caption{Dots represent the cumulative distribution of standardized logarithmic returns $|Ret|$ (i.e., the logarithmic returns $ret$ detrended with their mean and rescaled with their standard deviation). The positive and the negative tails were merged by using absolute returns. The solid line represents the cumulative distribution of a random variable drawn from a normal distribution.}
\label{cdf}
\end{figure}

Fig. \ref{cdf} shows the cumulative distribution of standardized logarithmic returns $Ret$, i.e., logarithmic returns $ret(h) = \log p(h) - \log p(h-1)$ (see Fig. \ref{ret}) detrended by their mean and rescaled by their standard deviation. For comparison, the solid line represents the cumulative distribution of the standard normal distribution $N(0,1)$. One observes a clear deviation from the Gaussian decay with approximate power law scaling in the tail.
A log-log regression of points which satisfy the condition $|Ret|>2$ gives the slope: $-3.69 \pm 0.02$.
This is not far from results empirically obtained for various financial prices for daily frequencies \cite{MS00,BP00}.

\begin{figure}
\begin{center}
\includegraphics[scale=0.7]{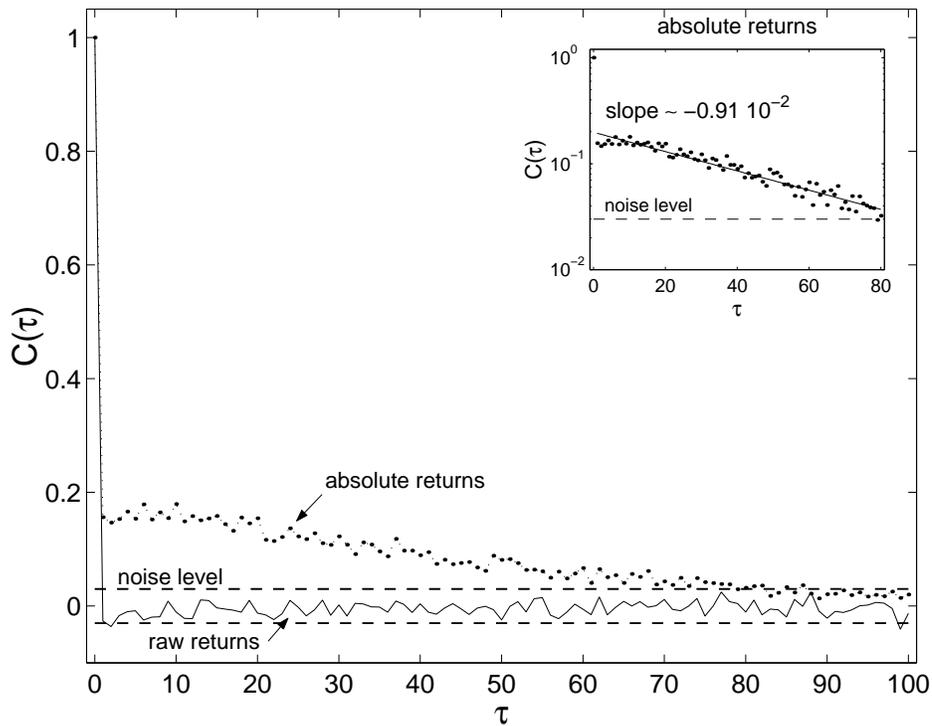}
\end{center}
\caption{The dotted line represents the autocorrelation of absolute returns; 
the solid line is the autocorrelation of raw returns. Noise levels are computed (see \protect{\cite{BP00}}) as $\pm 3/\sqrt{M}$ 
where $M$ is the length of the time series $(M = 10000)$. 
The inset exhibits the autocorrelation $C(\tau)$ for $\tau = 0,...,80$ in a semilog scale with a linear fit.}
\label{cov}
\end{figure}

In Fig. \ref{cov}, we present the autocorrelation $C(\tau)$ of the absolute returns $|ret|$ and of the raw returns $ret$ at different time lags $\tau$. While the autocorrelation of raw returns exhibits rapid decay, the autocorrelation of the absolute value of returns shows the presence of long-range correlations with a very slow exponential decay with exponent: $-(0.91 \pm 0.03) 10^{-2}$. Taking the absolute returns as a measure of volatility \cite{LGCMPS99} and considering the shape of Fig. \ref{ret}, the simulated time series exhibits the well known stylised fact of volatility clustering present in real-world markets.  \label{sec:experiments}

\section{Discussion and conclusions}
\label{sec:conclusions}
An interesting characteristic of the Genoa artificial market is its ability to exhibit the key stylised facts of financial time series (i.e., fat tails and volatility clustering) using simple trading rules in a realistic trading environment characterised by the finiteness of agent resources, order limit prices, and the creation and matching of demand and supply curves. 

There are some shortcomings of the model that we will address in future research. First, the volatility clustering effect is sensitive to model size. If the number of agents becomes very large, volatility clustering tends to disappear. Second, this model is still unable to correctly represent all the known stylised facts on price behaviour. The volatility exhibits an exponential decay, while empirical studies show a power law decay \cite{MS00,LGCMPS99}. However, it might be noted that a satisfactory microscopic explanation of the power law decay of volatility is still lacking \cite{Bouchaud01} and even the well known ARCH and GARCH models, based on correlating volatility at different time steps, exhibit a volatility exponential decay \cite{MS00}.

Note that, within the limits of our simulations, the scaling of volatility exhibits an exponent $< 0.5$. This might be related to the fact the Genoa market is a closed system with a mechanism of reversion to the mean in the price process. A coupling mechanism with an external cash generation process might change this finding. Actually, many real markets, e.g., some currency exchanges and commodities markets (copper, oil, electricity), exhibit a mean reverting behaviour \cite{Schwartz97}.

Finally, the Genoa market is a computational laboratory where many experiments can be performed. The simulator was conceived to evolve; it was implemented using object-oriented technology and extreme programming \cite{Beck99}. Using these techniques, it is possible to develop complex systems and to make substantial modifications very quickly, not jeopardizing quality. 

Future research will explore more sophisticated agent aggregation mechanisms and more intelligent trader behaviour. We will also add different kinds of securities and a book of orders; we also plan to extend simulation and theoretical analysis to a non-stationary environment, simulating the injection of cash in the system from an external simulated economy.

\section*{Acknowledgements}
The authors gratefully acknowledge useful comments by an anonymous referee. Gianaurelio Cuniberti carefully revised an early version of this paper. This work has been supported  by Fabbrica Servizi Telematici S.r.l, Cagliari, Italy.

\newpage

\newpage

\newpage

\end{document}